\documentclass[prd,superscriptaddress,twocolumn,aps,amsmath,amssymb,showpacs,nofootinbib]{revtex4-2}

\usepackage{amsmath}
\usepackage{graphicx,enumerate,xcolor,enumitem,xspace,braket,mathrsfs} 
\usepackage[colorlinks,citecolor=blue]{hyperref}
\usepackage{comment}

\usepackage{ntheorem}
\theoremheaderfont{\itshape}
\theorembodyfont{\upshape}

\newcommand{\nn}{\nonumber\\}
\newcommand{\tr}{\operatorname{Tr}}
\newcommand{\suma}{ \frac{1}{N} \sum_{a =1}^N}

\newcommand{\bfC}{\mathbf{C}}

\newcommand{\bbI}{\mathbb{I}}

\newcommand{\ie}{{\it i.e.}}

\newcommand{\rhs}{ r.h.s.\xspace}

\newcommand{\itbf}[1]{{\it \textbf{#1}}}
\newcommand{\customsection}[1]{\section{\bf #1}}

\usepackage[normalem]{ulem}

\begin{document}

{\hfill PITT-PACC-2321}
\title{Optimizing Fictitious States for Bell Inequality Violation in \\ Bipartite Qubit Systems with Applications to the $t\bar t$ System}

\author{Kun Cheng}
\email{chengkun@pku.edu.cn}
\affiliation{School of Physics, Peking University, Beijing, 100871, China} 
\affiliation{PITT PACC, Department of Physics and Astronomy,\\ University of Pittsburgh, 3941 O’Hara St., Pittsburgh, PA 15260, USA}

\author{Tao Han}
\email{than@pitt.edu}
\affiliation{PITT PACC, Department of Physics and Astronomy,\\ University of Pittsburgh, 3941 O’Hara St., Pittsburgh, PA 15260, USA}

\author{Matthew Low}
\email{mal431@pitt.edu}
\affiliation{PITT PACC, Department of Physics and Astronomy,\\ University of Pittsburgh, 3941 O’Hara St., Pittsburgh, PA 15260, USA}

\date{\today}

\begin{abstract}
There is a significant interest in testing quantum entanglement and Bell inequality violation in high-energy experiments. Since the analyses in high-energy experiments are performed with events statistically averaged over phase space, the states used to determine observables depend on the choice of coordinates through an event-dependent basis and are thus not genuine quantum states, but rather ``fictitious states.'' We find that the basis which diagonalizes the spin-spin correlations is optimal for constructing fictitious states to test the violation of Bell's inequality.  This result is applied directly to the bipartite qubit system of a top and anti-top produced at a hadron collider. We show that the beam axis is the optimal basis choice near the $t\bar t$ threshold production for measuring Bell inequality violation, while at high transverse momentum the basis that aligns along the momentum direction of the top is optimal.
\end{abstract}

\maketitle

\customsection{Introduction}

Quantum mechanics and the theory of relativity are the two pillars of modern physics.  Quantum entanglement and the violation of Bell's inequality~\cite{Bell:1964kc} are the unambiguous tests of quantum principles, and have been verified in numerous low-energy experiments.  Recently, there has been a significant interest in testing quantum entanglement and Bell inequality violation in high-energy experiments~\cite{Afik:2020onf}.  While it is important to test the principles of quantum mechanics deeply into the relativistic kinematic regime, certain aspects of high-energy experiments make these test non-trivial and even ambiguous. 

Quantum states, represented by density matrices, are the fundamental objects in quantum mechanics. The process to reconstruct the density matrix of a quantum state, known as quantum tomography, requires measurements of an ensemble of events. In high-energy experiments, each event specified by a multi-dimensional phase space point is a measurement of an underlying quantum state.  Collecting a large data sample with different phase space configurations is equivalent to a phase space integration weighted by a matrix element squared, which does not result in a quantum state when the average is done in event-dependent frames. Such an object has been labelled a ``fictitious state''~\cite{Afik:2022kwm}.

Fictitious states are a convex sum of quantum sub-states, but are not added in a way that corresponds to a quantum state.  Consequently, fictitious states lose many properties that are held by quantum states.  While observables computed using quantum states are basis-independent, observables computed using fictitious states are basis-dependent in general. For instance, the Bell inequality violation of a fictitious state becomes basis-dependent. 

In this work, we show that the fictitious states do maintain some properties of quantum states. In particular, the Bell inequality violation of a fictitious state implies the same for some quantum sub-state. This allows us to identify the basis that extremizes the Bell inequality violation of a fictitious state.
Using the top pair production process at the Large Hadron Collider (LHC) as an example, we present the optimal basis choice for constructing fictitious states which leads to the largest possible observation of Bell inequality violation in the top system at the LHC.

\vskip 1em
\customsection{Fictitious States and their Basis Dependence}

Consider a quantum state that is described by the density matrix $\rho$.  With $N$ measurements, we construct the quantum state $\overline{\rho}$ 
\begin{equation}
\label{eq:rhoaverage}
\overline{\rho} = \frac{1}{N} \sum_{a=1}^N \rho_a,
\end{equation}
where $\rho_a$ is a quantum sub-state from a single measurement $a$.

We first start with the simple example of a spin measurement on a single qubit, where we can decompose a quantum sub-state as
\begin{equation}
\label{eq:decomposition}
\rho_a = \frac{1}{2}\left( \mathbb{I}_2 + \sum_{i=1}^3 B_{a,i} \sigma_i \right).
\end{equation}
The reconstructed quantum state is an average over the $N$ measurements
\begin{align}
\label{eq:rhobardef}
\overline{\rho} &= \frac{1}{2}\left( \mathbb{I}_2 + \sum_{i=1}^3 \overline{B}_i \sigma_i \right) 
\quad {\rm with} \quad
\overline{B}_i 
= \frac{1}{N} \sum_{a=1}^N B_{a,i}.
\end{align}
The spin of the qubit is measured by $\langle \sigma_i \rangle$:
\begin{align}
\label{eq:spinofQubit}
\langle \sigma_i \rangle
&= \frac{1}{N} \sum_{a=1}^N \tr(\sigma_i \rho_a) 
= \frac{1}{2N} \sum_{a=1}^N  \tr \left( \sigma_i (\sum_{j=1}^3 B_{a,j} \sigma_j) \right) \nn
&  =  \frac{1}{2} \tr \left( \sigma_i (\sum_{j=1}^3 \overline{B}_j\sigma_j) \right) = \overline{B}_i  = \tr(\sigma_i \overline\rho) .
\end{align}
Next, consider reconstructing the state with a different measurement scheme where rather than using the same reference axis for each measurement $a$: $\sigma_i = \vec{\sigma} \cdot \hat{e}_i$, instead a different reference axis is used for each measurement $a$: $\sigma_{i,a} = \vec{\sigma} \cdot \hat{e}_{i,a}$.  We call this scheme using an \textit{event-dependent basis}, and the previous event-independent scheme using a \textit{fixed basis}.  They are related by an event-dependent rotation $R_{ij,a}$ such that $\hat{e}_{i,a} =\sum_j R^T_{ij,a} \hat{e}_{j}$.

Using an event-dependent basis, the basis for each measured quantum sub-state is different 
\begin{equation}
\label{eq:rhoqubitBprime}
\rho_{a} = \frac{1}{2}\left( \mathbb{I}_2 + \sum_{i=1}^3 B'_{a,i} \sigma_{i,a} \right).
\end{equation}
The corresponding reconstructed averaged state is 
\begin{equation}
\label{eq:fict}
\overline{\rho}_{\rm fic}
= \frac{1}{2}\left( \mathbb{I}_2
+ \sum_{i=1}^3 \overline{B}'_i \sigma_i
\right)
\quad {\rm with} \quad
\overline{B}'_i 
= \frac{1}{N} \sum_{a=1}^N B'_{a,i}.
\end{equation}
Under the parametrization of Eq.~\eqref{eq:rhoqubitBprime}, the measured value of the spin $\langle \sigma_i \rangle$ compares to the calculated value $\tr(\sigma_i \overline{\rho}_{\rm fic})$ as
\begin{align}
\langle \sigma_i \rangle
&= \frac{1}{N} \sum_{a=1}^N \tr(\sigma_i \rho_{a}) = \frac{1}{2N} \sum_{a=1}^N  \tr \left( \sigma_i (\sum_{j=1}^3 B'_{a,j} \sigma_{j,a}) \right) \nn
\label{eq:evtbyevt}
& = \frac{1}{N} \sum_{a=1}^N
\sum_{j=1}^3 \left( R_{ij,a} B'_{a,j} \right) \neq  \overline{B}'_i = \tr(\sigma_i \overline{\rho}_{\rm fic}),
\end{align}
where we have used $\tr(\sigma_i \sigma_{j,a} )/2 =R_{ij,a}$.  As such, the fictitious state reconstructed using the event-dependent basis does not yield the same result as the genuine quantum state.

Fictitious states, given by $\overline{\rho}_{\rm fic}$, are still a convex sum of quantum sub-states, but they are not added in a way that corresponds to a quantum state. This drawback is more pronounced  for other observables beyond spin.  For any space-time-dependent observable $\mathcal{O}$, the predicted result $\tr(\mathcal{O}\,\overline{\rho}_{\rm fic})$ would depend on the phase space configuration and thus on the choice of the Lorentz frame, as well as the coordinate axes.

A general remark is in order.  Eq.~\eqref{eq:rhoaverage} can always be rewritten as a sum of weighted events $\overline \rho \propto \sum_a \omega_a\rho_a$ where the weight $\omega_a$ denotes the production rate, or it can be changed from a summation to a phase space integral  $\overline\rho\propto\int \frac{\rm d\sigma}{\rm d\Omega} \rho_{\Omega} {\rm d\Omega}$.  These formulations are often more suitable for the quantum states produced at collider experiments.  All of our discussion still holds as long as $\overline\rho$ is a convex sum of quantum sub-states. Additionally, while the previous example is based on the spin measurement of a spin-1/2 particle, we emphasize that the preceding discussion applies to any fictitious state and its corresponding quantum state.

For a generic qubit state, the density matrix can be expressed as a $2\times 2$ matrix $\rho = \rho_{\alpha\beta} \ket{\alpha} \bra{\beta}$.  Hereafter, the summation of repeated indices on density matrices is assumed.  With respect to a different basis choice, the qubit state in the primed basis corresponds to an $SU(2)$ transformation $\ket{\beta'}=\ket{\alpha}U_{\alpha\beta'}$, which acts on the density matrix according to $\rho' = U^\dagger \rho U$. 

For spin-1/2 particles, the quantization basis choice is made by choosing reference axes $\hat{e}_i\ (i=1,2,3)$ to measure their spin.  Different reference axes are related by an $SO(3)$ rotation $\hat{e}_i' = \sum_j R^T_{ij} \hat{e}_j$. Consequently, the $2\times 2$ density matrix of a qubit expressed in the basis using $\hat e_i$ and the basis using $\hat e'_i$ are related by
\begin{equation}
\label{eq:rhoU}
    \rho^{\hat e'_i}_{\alpha'\beta'}= U^\dagger_{\alpha'\alpha}\rho^{\hat e_i}_{\alpha\beta}U_{\beta\beta'}.
\end{equation}
The $SO(3)$ rotation from $\hat e_i$ to $\hat e'_i$ is given by $R_{ij}=\tr(U^\dagger \sigma_i U \sigma_j)/2$, which is a generalization from the earlier example. 

Considering the state averaged over a large ensemble with a different basis $\hat{e}_{i,a}$ for each event $a$, 
\begin{equation}\label{eq:rhobarFictitious}
\overline\rho[{\hat e_{i,a}}] = \suma \rho_a^{\hat e_{i,a}}
= \suma U^\dagger_{a} \rho_a^{\hat e_{i}} U_{a},
\end{equation}
where $\rho[{\hat e_{i,a}}]$ denotes the reconstruction of the state using an event-dependent basis.  This relation recasts our previous conclusion in a general form.  Although the genuine quantum state averaged in two different -- but fixed -- bases are equal with respect to a unitary rotation, as shown in Eq.~\eqref{eq:rhoU}, the density matrix $\overline \rho[\hat e_{i,a}]$ reconstructed using an event-dependent basis is generally not equivalent to the genuine quantum state reconstructed using a fixed basis: $\overline \rho[\hat e_{i,a}] \neq U^\dagger \overline \rho^{\hat e_i} U$.  Therefore, the density matrix averaged using an event-dependent basis is a fictitious state.  Moreover, the density matrix averaged using two different event-dependent basis schemes are not equivalent, {\it i.e.}, $\overline \rho[\hat e_{i,a}] \ne U^\dagger \overline \rho[\hat e'_{i,a}] U$.  In other words, fictitious states are basis-dependent, which leaves room for an optimal choice of the basis for a given physical purpose. 

This formalism applies equally to two qubits -- the bipartite qubit system -- which is the simplest system that can exhibit entanglement and Bell inequality violation.  In collider experiments, this system is the most extensively studied via the production of a top quark and an anti-top quark~\cite{Afik:2020onf,Fabbrichesi:2021npl,Afik:2022kwm}.

A bipartite qubit system is constructed by a direct product of the two qubits, and its density matrix is a $4\times 4$ matrix $\rho_{\alpha\bar\alpha,\beta\bar\beta}$ expressed in the basis $\ket{\alpha}\otimes\ket{\bar\alpha}$.  Following the previous notation, the $4\times 4$ density matrix of a bipartite qubit system expressed in different bases are related by 
\begin{align}
\rho^{\hat e'_i}_{\beta'\bar\beta',\alpha'\bar\alpha'}&= U^\dagger_{\beta'\bar{\beta}',\beta\bar{\beta}}\rho^{\hat e_i}_{\beta\bar\beta,\alpha\bar{\alpha}}U_{\alpha\bar{\alpha},\alpha'\bar{\alpha}'},\\
U_{\alpha\bar\alpha,\alpha'\bar\alpha'}&\equiv U_{\alpha\alpha'}\otimes U_{\bar\alpha'\bar\alpha'}. \nonumber
\end{align}
In the following, we focus on optimizing the choice of fictitious state for the purpose of testing Bell inequality violation.

\vskip 1em
\customsection{The Bell Inequality Violation of Fictitious States}

To study the Bell inequality violation of a bipartite qubit system, it is convenient to parameterize the $4\times 4$ density matrix as
\begin{align}
\label{eq:rhoBiCij}
\rho=\frac{1}{4}&\bigg( \bbI_2\otimes \bbI_2 + \sum_{i=1}^3 B_i^+ \sigma_i\otimes \bbI_2 + \sum_{i=1}^3 B_i^- \bbI_2\otimes \sigma_i \nn
&+ \sum_{i,j=1}^3 C_{ij} \sigma_i\otimes \sigma_j \bigg).
\end{align}
The correlation matrix $\mathbf{C}$ -- written above in component form as $C_{ij}$ -- encodes the correlations relevant for Bell's inequality.  The Bell's inequality for bipartite qubit systems is the Clauser-Horne-Shimony-Holt inequality (CHSH)~\cite{Clauser:1969ny}, written as
\begin{align}
\label{eq:Bella1a2b1b2}
    \left| \Vec{a}_1\cdot \bfC\cdot (\vec{b}_1-\Vec{b}_2)+
    \Vec{a}_2\cdot \bfC\cdot (\vec{b}_1+\Vec{b}_2) \right| \leq 2
\end{align}
where $\vec a_{1,2}\ (\vec b_{1,2})$ are normalized directions to measure the spin of the first (second) qubit.  Classical theories, including theories with local hidden variables, satisfy this inequality.  The violation of Bell's inequality is a consequence of genuine quantum mechanical behavior.

The optimal choice of the four spin measurements in Eq.~\eqref{eq:Bella1a2b1b2} is known to be~\cite{HORODECKI1995340}
\begin{align}
\label{eq:Brho}
\mathscr{B}(\rho)&=\max_{\vec a_1, \vec a_2, \vec b_1, \vec b_2}
\left| \Vec{a}_1\cdot \bfC\cdot (\vec{b}_1-\Vec{b}_2)+
\Vec{a}_2\cdot \bfC\cdot (\vec{b}_1+\Vec{b}_2) \right|\nn
&= 2 \max_{i\neq j} \sqrt{\mu_i^2+\mu_j^2},
\end{align}
where $\mu_i^2$ are the eigenvalues of $\mathbf{C}^T \mathbf{C}$.  In the following, we assume that the correlation matrix $\bfC$ is symmetric, in which case $\mu_i$ are eigenvalues of $\mathbf{C}$. 

When we reparametrize the density matrix using another spin basis:   $\ket{\alpha}\to \ket{\beta}U_{\beta\alpha}$, the correlation matrix $\bfC$ transforms as an $SO(3)$ tensor: $\bfC\to R^T\bfC R$.  Therefore, for a genuine quantum state, $\mathscr{B}(\rho)$ is basis-independent.  For a fictitious state, however, its correlation matrix is obtained by averaging the correlation matrices of quantum sub-states using an event-dependent basis
\begin{align}
\label{eq:CbarFictitious}
 \overline\bfC[\hat e_{i,a}] =\suma  \bfC_a^{\hat e_{i,a}} = \suma R^T_a \bfC_a^{\hat e_i} R_a,
\end{align}
where $\bfC_a^{\hat e_i}$ is the correlation matrix of the sub-state $a$ expressed in a fixed basis and $R_a$ is an event-dependent rotation from a fixed basis $\hat e_i$ to the event-dependent basis $\hat e_{i,a}$.  Generally, the correlation matrices of two fictitious states averaged in different bases, $\overline\bfC[\hat e_{i,a}]$ and $\overline\bfC[\hat e_{i,a}']$, yield different results and different eigenvalues.  Thus, the violation of Bell's inequality is also basis-dependent, \ie, $\mathscr{B}(\rho[\hat e_{i,a}])\neq\mathscr{B}(\rho[\hat e_{i,a}'])$.

It is important to realize that establishing the Bell inequality violation of a fictitious state $\overline\rho_{\rm fic}$ implies the same for some quantum sub-state $\rho_a$. While this fits the naive expectation, we prove this in the following. 

Assume that all quantum sub-states $\rho_a$ satisfy the CHSH inequality, \ie, for any four spatial directions $\vec{a}_1,\vec{a}_2,\vec{b}_1$, and $\vec{b}_2$,
\begin{align}
\label{eq:CHSHsubstates}
 \Vec{a}_1\cdot \bfC_a \cdot (\vec{b}_1-\Vec{b}_2)+
 \Vec{a}_2\cdot \bfC_a\cdot (\vec{b}_1+\Vec{b}_2) \in [-2,2],
\end{align}
where $\mathbf{C}_a$ is the correlation matrix of the $a^{\rm th}$ sub-state expressed in an arbitrary basis.  Here, we omit the basis choice label for simplicity when referring to a general event-dependent basis such that $\bfC_a^{\hat e_{i,a}}$ is written as $\bfC_a$ and $\overline \bfC[\hat e_{i,a}]$ is written as $\overline\bfC$.  The sum of Eq.~\eqref{eq:CHSHsubstates} is convex and yields
\begin{align}
    &\suma
    \left[ \Vec{a}_1\cdot \bfC_a\cdot(\vec{b}_1-\Vec{b}_2)+ \Vec{a}_2\cdot \bfC_a\cdot(\vec{b}_1+\Vec{b}_2) \right]\nn
    =&~ \Vec{a}_1\cdot \overline \bfC \cdot(\vec{b}_1-\Vec{b}_2)+ \Vec{a}_2\cdot \overline \bfC \cdot(\vec{b}_1+\Vec{b}_2)  \in  [-2,2]. \nonumber 
\end{align}
This implies that a fictitious state constructed from quantum sub-states, all of which satisfy Bell's inequality, must also satisfy Bell's inequality.  Conversely, establishing Bell inequality violation in a fictitious state shows there must be a quantum sub-state that violates Bell's inequality.

\vskip 1em
\customsection{The Optimal Basis for Bell Inequality Violation}

Since the construction of a fictitious state is basis-dependent, one may hope to find an optimal basis for a specified physical application. 
In this work, we analyze the $q\bar q\to t\bar t$ process.  At a hadron collider $t\bar t$ is produced both via $q \bar q$ and $gg$.  Our analysis straightforwardly applies to $gg \to t\bar t$ also.
We show that the optimal basis to test Bell's inequality is the one that diagonalizes the spin correlation of each quantum sub-state, and we call it the \textit{diagonal basis} (see the Appendix for detail).

For the $q\bar{q}\to t\bar{t}$ process, the diagonal basis $\hat e_{i,a}^{\rm diag}$ can be obtained by a rotation from the \textit{helicity basis} $(\hat{r},\hat{n},\hat{k})$, where the axis $\hat{k}$ of the helicity basis is aligned with the top quark momentum direction $\hat{k}$ in the top pair center-of-mass frame, $\hat{n}$ is the direction normal to the scattering plane, and $\hat{r}=\hat{n}\times \hat{k}$.  The rotation to the diagonal basis is
\begin{subequations}
\begin{align}
    \hat{e}_1^{\,\rm diag}&= \hat{r}\cos\xi+\hat{k}\sin\xi, \\
    \hat{e}_2^{\,\rm diag}&= \hat{n},\\
    \hat{e}_3^{\,\rm diag}&= \hat{k}\cos\xi- \hat{r}\sin\xi,
\end{align}
\end{subequations}
with $\tan\xi=\tan\theta/\gamma$~\cite{Mahlon:1997uc}, where $\theta$ is the scattering angle in the top pair center-of-mass frame, $\gamma=\sqrt{\hat{s}}/(2m_t)$ is the boosted factor, and $\sqrt{\hat{s}}$ is the center-of-mass energy of top pair system.

\begin{figure}[tbh]
    \centering
    \hspace{-10pt}
    \includegraphics[width=.5\linewidth]{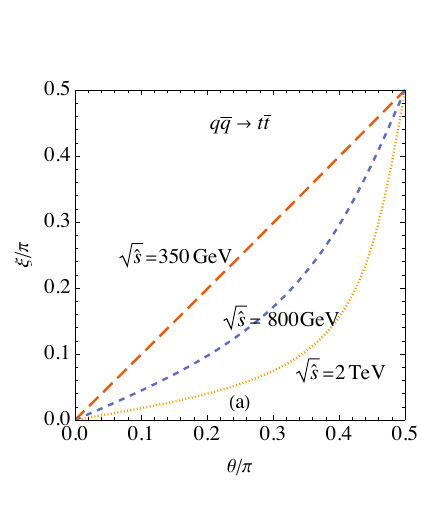} \hspace{-10pt}
    \includegraphics[width=.5\linewidth]{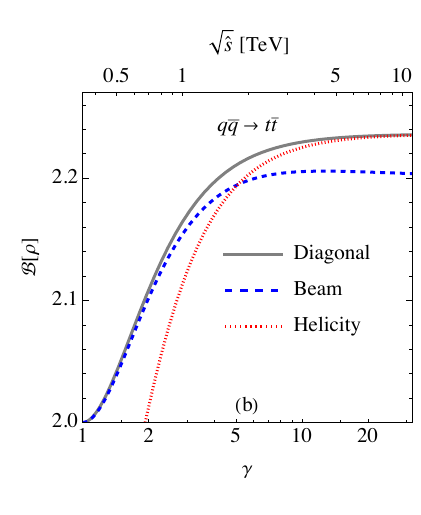}
    \caption{(a) The rotation angle $\xi$ from the helicity basis to the diagonal basis. (b) Bell inequality violation of $t\bar t$ produced from $q\bar q$ annihilation, measured in different event-dependent bases. }
    \label{fig:qqtt}
\end{figure}

Another useful basis is the \textit{beam basis} $(\hat{x},\hat{y},\hat{z})$, where the $z$-direction is fixed as the beam direction for all $t\bar t$ events, while the second direction $\hat{y}=\hat{n}$ is perpendicular to the scattering plane and $\hat{x}=\hat{z}\times\hat{y}$.
The correlation matrix in each of these three bases is appropriately averaged, using Eq.~\eqref{eq:CbarFictitious}, and we calculate the Bell inequality violation $\mathscr{B}[\overline{\rho}]$ for each basis.  To obtain the diagonal basis, the rotation angle ($\xi$) from the beam basis angle ($\theta$) is shown in Fig.~\ref{fig:qqtt}(a) for the $q\bar q \to t\bar t$ process.
The diagonal basis approaches the beam basis when the top pair is produced near threshold ($\sqrt{\hat s}=350$ GeV), while the diagonal basis approaches the helicity basis in the boosted region ($\sqrt{\hat s}=2$ TeV).
Rather than the commonly-used beam or helicity basis, a proper choice of spin axis according to different scattering angle and center-of-mass energy in Fig.~\ref{fig:qqtt}(a) provides a clear improvement on testing the violation of Bell's inequality. This is compared in Fig.~\ref{fig:qqtt}(b) for the Bell inequality violation of $t\bar t$ measured in the three bases.
While Fig.~\ref{fig:qqtt} is only for $q\bar q\to t\bar t$ process, it can be proved in general that the diagonal basis is the the optimal basis that maximizes the Bell inequality violation for any bipartite qubit system (see Appendix for the proof), and the diagonal basis can also be obtained analytically for realistic top pair production processes at the LHC~\cite{future}.
Moreover, our formalism of fictitious states in Eq.~\eqref{eq:rhobarFictitious} and Eq.~\eqref{eq:CbarFictitious} are also fit for event-dependent Lorentz frame choices, and the diagonal basis defined in the center-of-mass frame of the bipartite qubit system is find to be the optimal reference frame to construct fictitious states and test Bell inequality violation~\cite{future}.
Our finding is applicable to future development of quantum observables at colliders for a better understanding of the connection between quantum information and collider physics.

\vskip 1em
\customsection{Conclusions}

The density matrix of a quantum state is reconstructed from the expectation value of observables, which are measured from an average over a large event ensemble.  When the measurements are performed in an event-dependent frame, one reconstructs a fictitious state, rather than a genuine quantum state.  

We showed that, although the fictitious state loses most properties of a genuine quantum state and its construction is basis-dependent, an observation of Bell inequality violation based on an analysis of a fictitious state would still imply Bell inequality violation for a quantum sub-state.  Moreover, the basis-dependence of fictitious states leaves a freedom of a basis choice for its construction. We further demonstrated the optimal basis choice for the $q\bar q\to t\bar t$ process that maximizes the violation of Bell's inequality, and find significant improvement on the signal of Bell inequality violation using the optimal basis.

Our observations can be generalized to any qubit system, and have important consequences for quantum systems observables.  In particular, quantum mechanical systems, constructed from the spins of final state particles at colliders commonly use event-dependent bases.  For a given kinematic configuration, the axes of the diagonal basis can change which impacts how the spins should be measured.

Given the widespread applications of fictitious states at particle colliders, there are many possible future directions exploring other properties of fictitious states.  Beyond Bell inequality violation, it can also be shown that the diagonal basis yields the optimal fictitious state for entanglement~\cite{future}.  The orientation of the diagonal basis depends on the specific process in question and is potentially different in each application. 

In closing, the use of event-dependent bases in quantum experiments leads to the reconstruction of fictitious states rather than quantum states.  We have shown that fictitious states still have utility in demonstrating the presence of Bell inequality violation in a quantum system and that the optimal basis, which diagonalizes the spin correlation matrix, maximizes the violation of Bell's inequality.  We hope these observations about fictitious states will clarify their usage in high-energy experiments and establish the groundwork for future developments. 

\begin{acknowledgments}
\vskip 2em
This work was supported in part by the U.S.~Department of Energy under grant No.~DE-SC0007914 and in part by the Pitt PACC.  TH would like to thank the Aspen Center for Physics, where part of this work is complete, which is supported by the National Science Foundation (NSF) grant PHY-1607611.  ML is also supported by the National Science Foundation under grant No.~PHY-2112829. KC is supported in part by the National Science Foundation of China under grant No.~12235001.
\end{acknowledgments}

\appendix

\section{Proof of the Maximization}

Here, we provide a proof that the diagonal basis maximizes the violation of Bell's inequality.

The correlation matrix of each quantum sub-state in the diagonal basis is
\begin{equation}
\label{eq:Cdiagonal}
\bfC^{\rm diag}_a=
   \begin{pmatrix}
        \mu_{1,a} &0&0\\
        0&\mu_{2,a} &0\\
        0&0&\mu_{3,a} \\
    \end{pmatrix}.
\end{equation}
Without loss of generality, the eigenvalues are ordered as $\mu_{1,a}\geq \mu_{2,a} \geq \mu_{3,a}$.  The correlation matrix of the fictitious state averaged in the diagonal basis,
\begin{equation}
\label{eq:Cbardiagonal}
\overline \bfC^{\rm diag} = \suma \bfC^{\rm diag},
\end{equation}
is also diagonal, and three eigenvalues of $\overline{\bfC}^{\rm diag}$ are $ \bar \mu_i \equiv \suma \mu_{i,a} $.

Next, consider another arbitrary basis $\hat e_{i,a}$ that is related to the diagonal basis by a possibly event-dependent rotation $R_a$.  The correlation matrix of the fictitious state in this basis is 
\begin{equation}
\label{eq:Cbargeneral}
\overline \bfC=\suma \bfC_a= \suma R^T_a \bfC^{\rm diag}_a R_a.
\end{equation}
Here and after, we omit the basis choice label for simplicity when referring to a general event-dependent basis such that $\bfC_a^{\hat e_{i,a}}$ is written as $\bfC_a$ and $\overline \bfC[\hat e_{i,a}]$ is written as $\overline\bfC$.
The three eigenvalues of $\overline \bfC$, which we denote as $\bar c_i$, satisfy the following relations:
\begin{align}
\label{eq:trC1trC2}
    &\bar c_1 + \bar c_2 + \bar c_3=\bar \mu_1 + \bar \mu_2 + \bar \mu_3 = \tr(\overline \bfC), \\ 
\label{eq:mu1Ciimu3}
    &\bar \mu_1 \geq \bar c_{i} \geq \bar \mu_3 \quad (\text{for any~} i=1,2,3).
\end{align}
Eq.~\eqref{eq:trC1trC2} is a direct consequence from the trace of Eqs.~\eqref{eq:Cbardiagonal} and \eqref{eq:Cbargeneral}.  To prove Eq.~\eqref{eq:mu1Ciimu3}, which states that $\bar c_i$ are bounded by $\bar \mu_i$, we first denote the three eigenvectors of $\overline \bfC$ as $\hat{v}_i$.  The corresponding eigenvalue is then
\begin{align}
    \bar c_i=\hat{v}_i\cdot \overline \bfC \cdot \hat{v}_i
    =\suma \left( \hat{v}_i\cdot \bfC_a \cdot \hat{v}_i \right).
\end{align}
Applying Eq.~\eqref{eq:Cdiagonal} and $\bfC_a=R^T_a \bfC^{\rm diag}_a R_a$, we find that 
\begin{equation}
\hat{v}_i\cdot \bfC_a \cdot \hat{v}_i=\sum_{\ell=1}^3 \left|(R_a \cdot \hat v_i)_\ell\right|^2 \mu_{\ell,a} 
\end{equation}
is a convex sum of $\mu_{i,a}$.  This leads to
\begin{equation}
\label{eq:mu1kCkmu3k}
    \mu_{1,a} \geq \hat{v}_i\cdot \bfC_a \cdot \hat{v}_i \geq \mu_{3,a}.
\end{equation}
Therefore, Eq.~\eqref{eq:mu1Ciimu3} holds as a convex sum of Eq.~\eqref{eq:mu1kCkmu3k}.

The Bell inequality violation of a density matrix $\rho$ is given by the largest two eigenvalues of its spin correlation matrix.  With Eqs.~\eqref{eq:trC1trC2} and \eqref{eq:mu1Ciimu3}, we are ready to show that the diagonal basis maximizes  $\mathscr{B}(\overline\rho[\hat e_{i,a}])$, by proving that for any $i\neq j$, there exist $k\neq \ell$ satisfying
\begin{equation}
\label{eq:cicjlessthanmukmul}
\bar c_i^2+ \bar c_j^2 ~\leq~      \bar \mu_k^2+ \bar\mu_\ell^2,
\end{equation}
where $\bar\mu_i$ and $\bar c_i$ are the eigenvalues of the correlation matrix averaged in the diagonal basis and an arbitrary basis, respectively.

The relative signs of $\bar\mu_1$, $\bar\mu_2$ and $\bar \mu_3$ can be divided into the following three cases, and we prove Eq.~\eqref{eq:cicjlessthanmukmul} case by case.
\begin{enumerate}[label=(\alph*)]
    \item  $\bar\mu_1 \geq \bar\mu_2 \geq \bar\mu_3 \geq 0$,
    \item  $0\geq \bar\mu_1 \geq \bar\mu_2 \geq \bar\mu_3$,
    \item $\bar\mu_1 \geq 0 \geq \bar\mu_3$.
\end{enumerate}

\vspace{10pt}
\noindent\itbf{Case (a)}. In this case both $\bar c_i$ and $\bar c_j$ are positive. Without loss of generality, we assume $\bar c_i \geq \bar c_j$.

Though $\bar c_i$ is bounded by $\bar \mu_1$ and $\bar \mu_3$, it can be either larger or smaller than $\bar \mu_2$:

\vspace{5pt}
\noindent\textit{Case (a1)} $0\leq \bar c_i \leq \bar \mu_2$. We have $\bar c_i^2 + \bar c_j^2 \leq 2\bar\mu_2^2 \leq \bar\mu_1^2+\bar\mu_2^2 $.  

\vspace{5pt}
\noindent\textit{Case(a2)} $\bar \mu_2\leq \bar c_i \leq\bar \mu_1$.
First, $\bar \mu_2\leq \bar c_i \leq\bar \mu_1$ leads to
\begin{align}
\label{eq:mu2cimu1}
-\frac{\bar \mu_1-\bar \mu_2}{2}
\leq\frac{\bar \mu_1+\bar\mu_2}{2}-\bar c_i \leq & \frac{\bar\mu_1-\bar\mu_2}{2}.
\end{align}
Second, combining Eq.~\eqref{eq:trC1trC2} and Eq.~\eqref{eq:mu1Ciimu3}, we have $\bar c_i+\bar c_j \leq \bar \mu_1+ \bar \mu_2$ for $i\neq j$. Consequently,
\begin{equation}
\label{eq:maximumofCiiCjj2}
\bar c_i^2 + \bar c_j^2 \leq \bar c_i^2 + \left(\bar \mu_1 + \bar \mu_2 - \bar c_i\right)^2.
\end{equation}
Next, we define a function
\begin{align}
f(\Delta) &= \left(\frac{\bar \mu_1+\bar \mu_2}{2}+\Delta\right)^2+ \left(\frac{\bar \mu_1+\bar \mu_2}{2}-\Delta\right)^2 \nn
&=\frac{(\bar \mu_1+\bar \mu_2)^2}{2}+2\Delta^2,
\end{align}
which satisfies $f(\Delta_1)\leq f(\Delta_2)$ for $|\Delta_1|\leq |\Delta_2|$.  Then $\bar\mu_1^2+\bar \mu_2^2$ and the \rhs of Eq.~\eqref{eq:maximumofCiiCjj2} and can be rewritten using $f(\Delta)$:
\begin{align}
\label{eq:fmu}
 \bar c_i^2+ \left(\bar \mu_1+\bar \mu_2-\bar c_i\right)^2&=f\left(\frac{\bar\mu_1+\bar\mu_2}{2}-\bar c_i \right),\\
\label{eq:fmu2}
\bar\mu_1^2+\bar\mu_2^2&= f\left(\frac{\bar\mu_1-\bar\mu_2}{2}\right).
\end{align}
From Eq.~\eqref{eq:mu2cimu1}, the \rhs of Eq.~\eqref{eq:fmu} is smaller than the \rhs of Eq.~\eqref{eq:fmu2}, then  $\bar c_i^2 + \left(\bar \mu_1 + \bar \mu_2 - \bar c_i\right)^2 \leq \bar\mu_1^2+\bar\mu_2^2$.  In combination with Eq.~\eqref{eq:maximumofCiiCjj2}, we reach our conclusion, $\bar c_i^2+ \bar c_j^2 \leq \bar \mu_1^2 + \bar \mu_2^2 $.

\vspace{10pt}
\noindent\textit{\textbf{Case (b)}}. In this case both $\bar c_i$ and $\bar c_j$ are negative. Without loss of generality, we assume $|\bar c_i|\geq |\bar c_j|$,  \ie, $\bar c_i \leq \bar c_j$.
\vspace{5pt}

\noindent\textit{Case (b1)} $0\geq \bar c_i \geq \bar \mu_2$. We have $\bar c_i^2 + \bar c_j^2 \leq 2\bar\mu_2^2 \leq \bar\mu_2^2+\bar\mu_3^2 $.

\vspace{5pt}
\noindent\textit{Case (b2)} $\bar \mu_2 \geq \bar c_i \geq \bar\mu_3$. Firstly, from $\bar \mu_2 \geq \bar c_i \geq \bar\mu_3$, we have
\begin{equation}
\label{eq:mu2cimu3}
    -\frac{\bar \mu_2-\bar\mu_3}{2}\leq
    \frac{\bar\mu_2+\bar\mu_3}{2}-\bar c_i \leq
    \frac{\bar \mu_2-\bar\mu_3}{2}.
\end{equation}
Then, combining Eq.~\eqref{eq:trC1trC2} and Eq.~\eqref{eq:mu1Ciimu3}, we have
$\bar c_i+\bar c_j \geq \bar \mu_2+\bar \mu_3$ (note that now both $\bar c_i$ and $\bar c_j$ are negative), which leads to
\begin{equation}
\label{eq:maximumofCiiCjj2caseb}
    \bar c_i^2+ \bar c_j^2 \leq \bar c_i^2+ \left(\bar \mu_2+ \bar \mu_3-\bar c_i\right)^2.
\end{equation}
Similar to case (a), we define a function
\begin{align}
    g(\Delta)&= \left(\frac{\bar \mu_2+\bar \mu_3}{2}+\Delta\right)^2+ \left(\frac{\bar \mu_2+\bar \mu_3}{2}-\Delta\right)^2 \nn
    &=\frac{(\bar \mu_2+\bar \mu_3)^2}{2}+2\Delta^2,
\end{align}
that satisfies $g(\Delta_1)\leq g(\Delta_2)$ for $|\Delta_1|\leq |\Delta_2|$.
We rewrite $\bar\mu_2^2+\bar\mu_3^2$ and the r.h.s of Eq.~\eqref{eq:maximumofCiiCjj2caseb} as
\begin{align}
\label{eq:gmu}
    \bar c_i^2+ \left(\bar \mu_2+ \bar \mu_3-\bar c_i\right)^2 &= g\left(\frac{\bar\mu_2+\bar\mu_3}{2}-\bar c_i\right),   \\
    \label{eq:gmu2}
     \bar\mu_2^2+\bar\mu_3^2 &=g\left( \frac{\bar \mu_2-\bar\mu_3}{2}\right).
\end{align}
From Eq.~\eqref{eq:mu2cimu3}, the \rhs of Eq.~\eqref{eq:gmu} is smaller than the \rhs of Eq.~\eqref{eq:gmu2}, which implies that $\bar c_i^2 + \left(\bar \mu_2 + \bar \mu_3 - \bar c_i\right)^2 \leq \bar\mu_2^2+\bar\mu_3^2$.  In combination with Eq.~\eqref{eq:maximumofCiiCjj2caseb}, we reach our conclusion
$\bar c_i^2+ \bar c_j^2 \leq \bar \mu_2^2 + \bar \mu_3^2 $.

\vspace{10pt}
\noindent\textit{\textbf{Case (c)}}. In this case we have no constraints on the sign of $\bar c_i$ and $\bar c_j$, so we enumerate all the possibilities.
\vspace{5pt}

\noindent\textit{Case (c1) } both $\bar c_i$ and $\bar c_j$ are positive.
The proof in case~(a) only relies on the fact that both $\bar c_i$ and $\bar c_j$ are positive, therefore, we have $\bar c_i^2+ \bar c_j^2 \leq \bar\mu_1^2 + \bar\mu_2^2 $.

\vspace{5pt}
\noindent\textit{Case (c2) } both $\bar c_i$ and $\bar c_j$ are negative. 
The proof in case~(b) only relies on the fact that both $\bar c_i$ and $\bar c_j$ are negative, therefore, we have $\bar c_i^2+ \bar c_j^2 \leq \bar \mu_2^2 + \bar \mu_3^2 $.

\vspace{5pt}
\noindent\textit{Case (c3) } $\bar c_i$ is positive while $\bar c_j$ is negative. 
Then $|\bar c_i|<|\bar \mu_1|$ and $|\bar c_j|<|\bar \mu_3|$, and we have $\bar c_i^2+ \bar c_j^2 \leq \mu_1^2 + \mu_3^2 $.

\vspace{10pt}
In summary, we have proven that the fictitious state averaged in the diagonal basis gives the largest violation of Bell's inequality, $\mathscr B(\overline \rho^{\rm diag})\geq \mathscr B(\overline \rho)$.

\bibliographystyle{apsrev4-1}
\bibliography{refs}
\end{document}